\documentclass[aps,prl,twocolumn,superscriptaddress,amsmath,amssymb,showpacs]{revtex4}

\usepackage{amssymb}
\usepackage[dvips]{graphicx}
\usepackage{bm}
\usepackage{epsfig}
\usepackage[ansinew]{inputenc}

\begin{document}

\title{Fluctuations in a superconducting quantum critical point of multi-band metals}

\author{A. Ramires}
\affiliation{Centro Brasileiro de Pesquisas F\'{\i}sicas, Rua Dr. Xavier Sigaud 150, 22290-180, Rio de Janeiro, RJ, Brazil}

\author{M. A. Continentino}
\email{mucio@cbpf.br}
\affiliation{Centro Brasileiro de Pesquisas F\'{\i}sicas, Rua Dr. Xavier Sigaud 150, 22290-180, Rio de Janeiro, RJ, Brazil}

\date{\today}

\begin{abstract}
In multi-band metals quasi-particles arising from different
atomic orbitals coexist at a common Fermi surface. Superconductivity in
these materials may appear due to interactions within a band (intra-band) or
among the distinct metallic bands (inter-band). 
Here we consider the suppression of superconductivity in the intra-band case due to hybridization. The fluctuations at the superconducting quantum critical point (SQCP) are obtained calculating the response of the
system to a fictitious space and time dependent field, which couples to the
superconducting order parameter.  The appearance of superconductivity is
related to the divergence of a generalized susceptibility. For a single
band superconductor this coincides with the \textit{Thouless criterion}.
For fixed chemical potential and large hybridization, the superconducting state has many features in common with breached pair superconductivity with unpaired electrons at the Fermi surface. The T=0 phase transition from the superconductor to the normal state is in the universality class of the density-driven Bose-Einstein condensation. For fixed number of particles and in the strong coupling limit, the system still has an instability to the normal sate with increasing hybridization.
\end{abstract}

\maketitle


\section{Introduction}

In heavy fermion materials superconductivity can be suppressed in different
ways. Most commonly, this is accomplished  by an external magnetic field, applied
pressure or doping %
\cite{ramos, leticie, loh, bianchi, mgb2, chanclog, demuer, flouquet}, but the
point in the phase diagram where the critical temperature $T_{c}$
vanishes as a function of the external parameters is not necessarily
associated with a SQCP. For example,  in the case of
superconductivity induced by antiferromagnetic fluctuations close to
an antiferromagnetic quantum critical point (AFQCP)\cite{steglich}, as the system
moves away from the AFQCP, these fluctuations change from
attractive to repulsive and superconductivity just fades away \cite%
{muciojpsj}.

Recently, we proposed a new mechanism for destroying superconductivity which
is specific for multi-band materials as heavy fermions \cite{Padilha, ramos}. It
occurs because the narrow band of quasi-particles which form the Cooper
pairs is hybridized with a band of normal electrons. As hybridization
increases by pressure or doping it turns out that a minimum attractive
interaction is required to produce a superconducting ground state. This
mechanism of suppression of superconductivity has been demonstrated using a
mean-field approach. Starting in the superconducting phase as hybridization
is turned on, $T_c$ was shown to vanish at a critical value of the mixing ($%
V_c$). However, the mean-field approach does not include fluctuations.
Although the normal state for $V>V_c$ is metallic, there are no critical
fluctuations associated with the onset of superconductivity in this approximation. Here we go
beyond the mean-field approximation to include fluctuations. We start in
the normal phase, where either $V>V_c$ or the attractive interaction is not
sufficiently strong to produce a superconducting ground state. As
hybridization decreases, or U increases, the system has an instability to a
superconducting ground state. This is shown using a new approach \cite{Ramires}  based on a
perturbation theory for retarded and advanced Green's functions \cite{tyablikov}.
For the single band case our results reduce to the well known
\textit{Thouless criterion} \cite{Thouless} for superconductivity. We relate
the appearance of superconductivity in the multi-band metal to the
divergence of a generalized susceptibility, very much like the Stoner
criterion for the appearance of ferromagnetism.

We calculate the response of the system to a frequency and wave-vector
dependent fictitious external field which couples to the superconducting
order parameter. For simplicity we consider here the case of s-wave
superconductivity. Using an RPA approximation, equivalent to summing an
infinite series of bubble diagrams, we obtain a generalized susceptibility $%
\chi(q, \omega)$. At zero temperature, the static and homogeneous part of
the susceptibility diverges at exactly the critical mean-field value of
hybridization, which destroys superconductivity. This divergence implies
that even in zero \textit{field} the system can have a finite
superconducting order parameter. The condition for the superconducting
instability can be expressed in the form of a Stoner-like criterion, $1-U
\chi_0(V)=0$. In the case with a fixed chemical potential this determines either a
critical value of the attractive interaction $U_c (V)$ above which the
system is superconductor or a critical hybridization $V_c(U)$ below which
superconductivity sets in.

The approach developed here allows to obtain the nature of the fluctuations close to the
SQCP.  We find that at $U_c(V)$ (or $V_c(U)$) and for low frequencies and small
wave-vectors, the generalized
frequency and wave-vector susceptibility has poles at real frequencies $%
\omega=\mathcal{D} q^2$.  It turns out that the SQCP is in the universality class of the Bose-Einstein condensation \cite{muciobe}. From the nature
of the quantum critical fluctuations, we obtain the dynamic critical
exponent $z$ using Hertz approach to quantum phase transitions \cite{hertz}.
This allows us to obtain the thermodynamic  properties when the
system is close to the SQCP \cite{mac}. Our theory is equivalent to a quantum Gaussian
approach and since the dynamic exponent turns out to be $z=2$, it yields the
correct description of the SQCP for dimensions $d \ge 2$. In our treatment the system is metallic in the non-superconducting side of the phase diagram. This arises due to quasi-particles which remain unpaired close to the Fermi surface. This is
different from the predictions of disordered induced SQCP \cite{dsqcp}, where
the normal state is insulating at $T=0$ due to the presence of a gap.
In the case of a single band and at zero temperature, the Thouless criterion
for BCS superconductivity implies a superconducting ground state for any
finite value of $U$.

It is interesting at this point to comment on the differences between the intra and inter-band problems.  We have learned from the mean-field approach \cite{Padilha} that in the former case, superconductivity is destroyed continuously through a SQCP as hybridization is increased. On the other hand for inter-band superconductivity, as hybridization increases superconductivity is suppressed discontinuously at zero temperature. This occurs at a first order quantum phase transition \cite{Padilha,first} with phase separation and metastable regions. For intra-band attractive interactions as one approaches the superconducting state from the normal phase, the instability of the normal metal is for a $q=0$ BCS state and occurs at the same value of hybridization for which the homogeneous BCS state is destroyed.
For the inter-band problem, the abrupt disappearance of superconductivity suggests to look for other forms of superconductivity which can remain stable in the presence of a large Fermi wave-vectors mismatch. For inter-band attractive interactions, the normal metal becomes unstable to an inhomogeneous $q\ne 0$ superconducting state for a value of the Fermi wave-vectors mismatch larger than that for which superconductivity is destroyed through a first order transition \cite{chanclog}. In a previous work, using the same approach proposed here, we could determine the universality class of the quantum phase transition from the normal to a pair density wave phase as the mismatch of the bands with inter-band interaction increases \cite{Ramires}.

The study of the crossover from weak coupling to strong coupling has
attracted considerable attention after the discovery of the
high-temperature superconductors. More recently, new experiments in
cold atom systems have further raised the interest in this problem
\cite{Sheehy}. Introducing the concept of scattering length,
$1/k_Fa_s$, \cite{SadeMelo2} we are able to analyze the case with
fixed number of particles and the crossover  between the weak and
strong coupling for the two-band system with hybridization. This is
necessary since the values of $U$ for which there is a critical
value of hybridization are in the strong coupling limit. In this
case the gap equation must be solved simultaneously with the number
equation and a trivia $(\mu, \Delta, 1/k_Fa_s)$ can be obtained
numerically. We find that the critical value of the hybridization
increases with the increase in the coupling strength tending to a
saturation value.

\section{The Model}

We start with the following Hamiltonian describing a two-band system \cite%
{suhl}, with a hybridization between them and an attractive interaction
in one of the bands \cite{japi, Padilha},
\begin{eqnarray}\label{Hamiltoniano0}
\mathcal{H}_0=\sum_{i,j,\sigma}t^a_{ij}a_{i\sigma}^{\dagger}a_{j \sigma}
+\sum_{i,j,\sigma}t^b_{ij}b_{i\sigma}^{\dagger}b_{j \sigma}-\frac{U}{2}%
\sum_{i,\sigma}n^a_{i\sigma}n^a_{i
-\sigma}\nonumber\\ +\sum_{i,j,\sigma}V_{ij}(b_{i\sigma}^{\dagger}a_{j
\sigma}+a_{i\sigma}^{\dagger}b_{j \sigma}),
\end{eqnarray}
where $a_{i \sigma}^{\dagger}$, $a_{j \sigma}$, $b_{i\sigma}^{\dagger}$, $%
b_{j \sigma}$ create and destroy electrons in the narrow $a$-band and the
wide $b$-band, respectively. The attractive interaction $U$ whose origin we
do not specify acts only between the electrons in the narrow $a$-band. The
hybridization $V$ mixes the electrons of different bands and can be
controlled by external pressure. In order to calculate the superconducting
response of this system, we proceed with the approach proposed in Ref. \cite{Ramires},
introducing a wave-vector and frequency dependent fictitious field that couples to the superconducting order parameter \cite%
{Cote},
\begin{equation}  \label{Hamiltoniano1}
\mathcal{H}_1= - g\sum_{i}e^{-i\mathbf{q} \cdot
\mathbf{r}_\mathbf{i}}e^{i \omega_0
t}(a_{i\uparrow}^{\dagger}a_{i\downarrow}^{\dagger}
+a_{i\uparrow}a_{i\downarrow}),
\end{equation}
where the frequency $\omega_{0}$ has a small positive imaginary part
to guarantee the adiabatic switching on of the field. The response
of the system to the fictitious field, will be obtained using
perturbation theory for the retarded and advanced Green's
functions \cite{tyablikov}. We start in the normal phase where the
superconducting order parameter is zero in the absence of the
\textit{external field} $g$. We split the Green's functions, normal
and anomalous in two contributions. The first of order zero and the
second of first order in the field $g$. For the anomalous Green's functions, we write,
\[
{\ll a_{i\sigma }^{\dagger }|a_{j\sigma ^{\prime }}^{\dagger }\gg }%
\rightarrow {\ll a_{i\sigma }^{\dagger }|a_{j\sigma ^{\prime }}^{\dagger
}\gg }^{(0)}+{\ll a_{i\sigma }^{\dagger }|a_{j\sigma ^{\prime }}^{\dagger
}\gg }^{(1)}.
\]

In the normal phase and in the absence of the fictitious field, the relevant
zero order Green's functions can be easily calculated \cite{Padilha}. They
are given by,%
\begin{eqnarray}\label{Gaa}
\mathcal{G}^{aa}(\omega, k)&=&{\ll a_{k\sigma }|a_{k\sigma^{\prime} }^{\dagger }\gg }^{(0)}_{\omega}\nonumber\\&=&\frac{(\omega -\epsilon _{k}^{b})\delta
_{k,k^{\prime }}\delta_{\sigma,\sigma^{\prime}}}{2\pi \lbrack (\omega
-\epsilon _{k}^{a})(\omega -\epsilon _{k}^{b})-V_{k}^{2}]},
\end{eqnarray}
\[
{\ll b_{k\sigma }|b_{k^{\prime }\sigma ^{\prime }}^{\dagger }\gg }%
^{(0)}_{\omega}=\frac{(\omega -\epsilon _{k}^{a})\delta _{k,k^{\prime
}}\delta _{\sigma ,\sigma ^{\prime }}}{2\pi \lbrack (\omega -\epsilon
_{k}^{a})(\omega -\epsilon _{k}^{b})-V_{k}^{2}]},
\]%
\[
{\ll a_{k\sigma }|b_{k^{\prime }\sigma ^{\prime }}^{\dagger }\gg }%
^{(0)}_{\omega}=\frac{V_{k}\delta _{k,k^{\prime }}\delta _{\sigma ,\sigma
^{\prime }}}{2\pi \lbrack (\omega -\epsilon _{k}^{a})(\omega -\epsilon
_{k}^{b})-V_{k}^{2}]},
\]%
\[
{\ll a_{k\sigma }^{\dagger }|a_{k^{\prime }\sigma ^{\prime }}^{\dagger }\gg }%
^{(0)}_{\omega}=0,
\]%
\[
{\ll b_{k\sigma }^{\dagger }|b_{k^{\prime }\sigma ^{\prime }}^{\dagger }\gg }%
^{(0)}_{\omega}=0,
\]
and
\[
{\ll a_{k\sigma }^{\dagger }|b_{k^{\prime }\sigma ^{\prime }}^{\dagger }\gg }%
^{(0)}_{\omega}={\ll b_{k\sigma }^{\dagger }|a_{k^{\prime }\sigma ^{\prime }}^{\dagger }\gg }%
^{(0)}_{\omega}=0.
\]

The first order propagators should be obtained more carefully due to the time dependence of the \textit{external field}. Let us consider the equation of motion for the first order anomalous Green's
function in the normal phase,
{\small
\begin{eqnarray*}
&i\frac{\partial}{\partial t}{\ll
a^{\dagger}_{i\sigma}(t)|a^{\dagger}_{j\sigma^{\prime}}(t^{\prime})\gg}%
^{(1)} = -\sum_{l}t^a_{il}{\ll a_{l\sigma }^{\dagger }(t)|
a_{j\sigma^{\prime}}^{\dagger}(t^{\prime} )\gg}^{(1)} \\
& +U{\ll a^{\dagger}_{i\sigma}(t)a^{\dagger}_{i-\sigma}(t)a_{i -\sigma}(t)|
a_{j\sigma^{\prime}}^{\dagger}(t^{\prime}) \gg}^{(0)} \\
& - \sum_{l} V_{il} {\ll b^{\dagger}_{l \sigma}(t) | a_{j
\sigma^{\prime}}^{\dagger} (t^{\prime} ) \gg}^{(1)} \\ & + g
\widetilde{\sigma} e^{-i \mathbf{q.r}_{i}} e^{i \omega_{0} t} {\ll
a_{i-\sigma}(t) | a_{j \sigma^{\prime}}^{\dagger}(t^{\prime})
\gg}^{(0)},
\end{eqnarray*}
}where $\widetilde{\sigma}=\pm1$ para $\sigma=\uparrow,\downarrow$.

We decouple the higher order Green's function in the spirit of the random
phase approximation (RPA) to obtain
{\small
\begin{eqnarray*}
&i\frac{\partial}{\partial t}{\ll
a^{\dagger}_{i\sigma}(t)|a^{\dagger}_{j\sigma^{\prime}}(t^{\prime})\gg}%
^{(1)} = -\sum_{l}t^a_{il}{\ll a_{l\sigma }^{\dagger }(t)|
a_{j\sigma^{\prime}}^{\dagger}(t^{\prime} )\gg}^{(1)} \\
&+U\delta \Delta_{i \sigma}(t){\ll a_{i -\sigma}(t)|
a_{j\sigma^{\prime}}^{\dagger}(t^{\prime} )\gg}^{(0)} \\
&-\sum_{l} V_{il} {\ll b^{\dagger}_{l \sigma}(t) | a_{j
\sigma^{\prime}}^{\dagger} (t^{\prime} ) \gg}^{(1)} \\ & + g
\widetilde{\sigma} e^{-i \mathbf{q.r}_{i}} e^{i \omega_{0} t} {\ll
a_{i-\sigma}(t) | a_{j \sigma^{\prime}}^{\dagger}(t^{\prime})
\gg}^{(0)},
\end{eqnarray*}
}
where in the adiabatic approximation,
\begin{equation}
\delta \Delta_{i \sigma}(t)=\delta \Delta_{\sigma}^{aa} e^{-i\mathbf{q.r_{i}}%
}e^{i \omega_{0} t}.
\end{equation}

Next, we perform a Fourier transformation only in the time variable $t$, using that
$a^{\dagger}_{i\sigma}(t)=\int d \omega a^{\dagger}_{i\sigma}(\omega) e^{-i
\omega t}$ and the same for $b^{\dagger}_{i \sigma}(t)$. Notice that the
zero order propagators are time translation invariant and depend only in the
time difference $(t-t^{\prime})$. The first order propagators however are
functions of both $t$ and $t^{\prime}$. We can also Fourier transform in
space, using that $a^{\dagger}_{i\sigma}(t)=\sum_{\mathbf{k}} a^{\dagger}_{%
\mathbf{k} \sigma}(t) e^{i \mathbf{k.r}_{i}}$. We get
{\scriptsize
\begin{eqnarray*}
&(\omega+\varepsilon_k^a){\ll\! a^{\dagger}_{k
\sigma}|a^{\dagger}_{k^{\prime}\sigma^{\prime}}(t^{\prime})\!\gg}%
_{\omega}^{(1)} = U\delta \Delta^{aa}_{\sigma} {\ll\! a_{(\mathbf{k-q})
-\sigma}| a_{k^{\prime}\sigma^{\prime}}^{\dagger}(t^{\prime} )\!\gg}_{\omega
+\omega_0}^{(0)} \\
&-V_k {\ll\! b^{\dagger}_{k
\sigma}|a_{k^{\prime}\sigma^{\prime}}^{\dagger}(t^{\prime} )\!\gg}%
_{\omega}^{(1)}+g\widetilde{\sigma} {\ll\! a_{(\mathbf{k-q}) -\sigma}|
a_{k^{\prime}\sigma^{\prime}}^{\dagger}(t^{\prime} )\!\gg}_{\omega
+\omega_0}^{(0)}.
\end{eqnarray*}
}

Spatial translation invariance is lost due to the spatial dependence of the
\textit{external field}. Now, we go on to write the equations of motion for
the new generated Green's functions. Proceeding like above we arrive at the
following system of equations,
{\footnotesize
\begin{eqnarray}\label{mov1}
&(\omega +\epsilon _{k}^{a}){\ll \! a_{k\sigma }^{\dagger }|a_{k^{\prime
}-\sigma }^{\dagger }\!\gg }_{\omega }^{(1)}=U\delta \Delta _{\sigma }^{aa}{%
\ll \! a_{(k-q)-\sigma }|a_{k^{\prime }-\sigma }^{\dagger }\!\gg }_{\omega
+\omega _{0}}^{(0)}  \nonumber \\ &
-V_{k}{\ll\! b_{k\sigma }^{\dagger }|a_{k^{\prime }-\sigma }^{\dagger }\!\gg }%
_{\omega }^{(1)}+g\widetilde{\sigma }{\ll\! a_{(k-q)-\sigma
}|a_{k^{\prime
}-\sigma }^{\dagger }\!\gg }_{\omega +\omega _{0}}^{(0)},\label{sys1}
\end{eqnarray}
\begin{eqnarray}\label{mov2}
&(\omega -\epsilon _{k}^{a}){\ll\! a_{k-\sigma }|a_{k-\sigma }^{\dagger }\!\gg }%
_{\omega }^{(1)}=\!-U\delta \Delta _{\sigma }^{aa}{\ll\! a_{(k-q)\sigma
}^{\dagger }|a_{k^{\prime }-\sigma }^{\dagger }\!\gg }_{\omega +\omega
_{0}}^{(0)}  \nonumber \\
&\!\!\!\!\!+V_{k}{\ll\! b_{k-\sigma }|a_{k-\sigma }^{\dagger }\!\gg }_{\omega }^{(1)}+g%
\widetilde{\sigma }{\ll\! a_{(k-q)\sigma }^{\dagger }|a_{k^{\prime }-\sigma
}^{\dagger }\!\gg }_{\omega +\omega _{0}}^{(0)},
\end{eqnarray}
}
{\small
\begin{eqnarray}\label{mov3}
(\omega +\epsilon _{k}^{b}){\ll b_{k\sigma }^{\dagger }|a_{k-\sigma
}^{\dagger }\gg }_{\omega }^{(1)}=-V_{k}{\ll a_{k\sigma }^{\dagger
}|a_{k^{\prime }-\sigma }^{\dagger }\gg }_{\omega }^{(1)},\label{hybrid}
\end{eqnarray}
\begin{eqnarray}\label{mov4}
(\omega -\epsilon _{k}^{b}){\ll b_{k-\sigma }|a_{k^{\prime }-\sigma
}^{\dagger }\gg }_{\omega }^{(1)}=V_{k}{\ll a_{k-\sigma }|a_{k^{\prime
}-\sigma }^{\dagger }\gg }_{\omega }^{(1)}.\label{sys4}
\end{eqnarray}
}

We notice from Eq. \ref{hybrid} that inter-band pairing can be induced by
hybridization even in the absence of inter-band attractive interactions. In
the case of purely inter-band interactions, the normal-BCS superconductor
transition is discontinuous and phase separation occurs \cite{sarma, caldas}.
Here, however, these inter-band anomalous correlations induced by
hybridization do not affect the nature of the transition \cite{Padilha}.

The
system of equations \ref{sys1}-\ref{sys4} can be easily solved. In
particular we get for the anomalous Green's function,
\begin{eqnarray}
{\ll a_{k\sigma }^{\dagger }|a_{(k-q)-\sigma }^{\dagger }(t^{\prime})\!\! \gg }%
_{\omega }^{(1)}\!=\!\frac{(\omega +\epsilon _{k}^{b})\left[ U\delta
\Delta ^{aa}+g\widetilde{\sigma }\right]}{[(\omega +\epsilon
_{k}^{a})(\omega +\epsilon _{k}^{b})-V_{k}^{2}]} \nonumber\\ \times{\ll a_{(k-q)-\sigma
}|a_{(k-q)-\sigma }^{\dagger}\!\!\gg }_{\omega +\omega _{0}}^{(0)}.
\end{eqnarray}

Notice that the index $(t^{\prime})$ becomes redundant since the zero order
Green's functions are time translation invariants. The anomalous first order
propagator can finally be written as,
\begin{eqnarray}
&{\ll a_{k\sigma }^{\dagger }|a_{(k-q)-\sigma }^{\dagger }\gg }_{\omega
}^{(1)}=\frac{1}{2\pi }\frac{(\omega +\epsilon _{k}^{b})}{[(\omega +\epsilon
_{k}^{a})(\omega +\epsilon _{k}^{b})-V_{k}^{2}]}\times  \nonumber \\
&\frac{(\omega +\omega _{0}-\epsilon _{k-q}^{b})}{[(\omega +\omega
_{0}-\epsilon _{k-q}^{a})(\omega +\omega _{0}-\epsilon
_{k-q}^{b})-V_{k-q}^{2}]}\left[ U\delta \Delta ^{aa}+g\widetilde{\sigma}%
\right].
\end{eqnarray}

Then, the anomalous first order Green's function can be completely
determined in terms of the zero order equilibrium Green's functions
obtained previously. This allows to use the fluctuation-dissipation theorem
 \cite{tyablikov} to obtain the correlation function $\delta \Delta ^{aa}$
from the above anomalous Green's functions,
\[
\delta \Delta ^{aa}=\sum_{k}F_{\omega }\left\{ {\ll a_{k\sigma }^{\dagger
}|a_{k^{\prime }-\sigma }^{\dagger }\gg }_{\omega }^{(1)}\right\},
\]%
where $F_{\omega }\{G(\omega )\}=-\int d\omega f(\omega )[G(\omega
+i\epsilon )-G(\omega -i\epsilon )]$ is the statistical average of the
discontinuity of the Green's functions $G(\omega )$ on the real axis \cite%
{tyablikov}. The function $f(\omega )$ is the Fermi-Dirac distribution.
Defining the susceptibility (see Eq.\ref{Gaa}) as,
\begin{equation}  \label{suscep}
\chi_{0}^{aa} (q,\omega _{0})=-2\pi\sum_{k}F_{\omega }\{\mathcal{G}%
^{aa}(-\omega ,k)\mathcal{G}^{aa}(\omega +\omega _{0},k-q)\},
\end{equation}
we can write the response to the fictitious field in the form,
\begin{equation}  \label{suscepg}
\delta \Delta ^{aa}=\widetilde{\sigma}\frac{\chi_{0}^{aa} (q,\omega
_{0})}{1-U\chi_{0}^{aa} (q,\omega _{0})}g.
\end{equation}

Then, the change of the order parameter due to the fictitious field is given
by a generalized susceptibility $\chi^{aa}(q, \omega_{0})$, in the form $%
\delta \Delta^{aa}=\chi^{aa}(q,\omega_{0})g$. As parameters of the
multi-band system in the normal metallic phase, such as, the hybridization or the
strength of the
attractive interaction change, eventually the condition $1-U\chi_{0}^{aa}(q,%
\omega_{0}=0)=0$ is attained. This signals  an instability of the system to a
superconducting phase since, even in the absence of the field $g$,
the order parameter may be finite. So, the appearance of
superconductivity is related to the divergence of a generalized
susceptibility. If the above criterion is first satisfied for $q=0$,
the instability is towards a homogeneous superconducting phase, if it
occurs first for a finite $q=q_{0}$, an inhomogeneous
superconducting FFLO-like phase \cite{fflo} sets in. This is the case
in general for inter-band interactions in the presence of a mismatch
of the Fermi-wave-vectors of the non-hybridized bands as is shown in Ref. \cite{Ramires}. For intra-band interactions, the instability is towards a homogeneous superconducting
phase when the condition $1-U\chi_{0}^{aa}(q=0,\omega_{0}=0)=0$ is
fulfilled \cite{Padilha}.

\subsection{The generalized susceptibility}

\begin{figure}[h]
\begin{center}
\includegraphics[width=0.9\linewidth,
keepaspectratio]{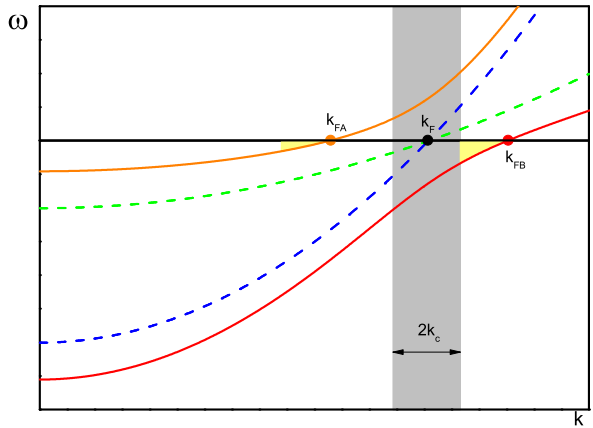}
\end{center}
\caption{(Color online) Dispersion relations for homothetic bands. The dashed lines represent the original unhybridized bands and the full lines the hybrid bands. For $V > V_{BCS}$ (see text), the region delimited by the cut-off vector $k_C$ around the Fermi wave-vector $k_F$ of the original bands  does not include the Fermi wave-vectors $k_{FA}$ and $k_{FB}$ of  the hybrid bands, as shown in the figure.} \label{fig1}
\end{figure}

Making a change of variables, and writing in a more symmetric form, the
frequency and wave-vector dependent susceptibility of Eq. \ref{suscep} is
obtained as,
\begin{widetext}
{\footnotesize
\begin{eqnarray}  \label{ChiSim}
\chi_{0}^{aa}(2q,\omega_0) =&
-\sum_{k}\frac{1}{(\omega^a_{k+q} + \omega^a_{k-q} - \omega_0)} \Bigg[%
\lbrack 1 - f(\omega^a_{k+q})]\frac{(\omega^a_{k+q} -
\epsilon_{k+q}^b)(\omega^a_{k+q} + \epsilon_{k-q}^b - \omega_0)}{%
(\omega^b_{k+q} - \omega^a_{k+q})(\omega^a_{k+q} + \omega^b_{k-q} - \omega_0)%
} + f(\omega^a_{k-q}) \frac{(\omega^a_{k-q} - \epsilon_{k-q}^b)
(\omega^a_{k-q} + \epsilon_{k+q}^b - \omega_0)}{( \omega^a_{k-q}
-\omega^b_{k-q})(\omega^b_{k+q} + \omega^a_{k-q} - \omega_0 )} \Bigg]
\nonumber\\&-\sum_{k} \frac{1}{(\omega^b_{k+q} + \omega^b_{k-q} - \omega_0)} \Bigg[ %
\lbrack 1-f(\omega^b_{k+q})]\frac{(\omega^b_{k+q} -
\epsilon_{k+q}^b)(\omega^b_{k+q} + \epsilon_{k-q}^b - \omega_0)}{%
(\omega^a_{k+q} - \omega^b_{k+q})(\omega^b_{k+q} + \omega^a_{k-q} - \omega_0)%
} + f(\omega^b_{k-q}) \frac{(\omega^b_{k-q} - \epsilon_{k-q}^b)
(\omega^b_{k-q} + \epsilon_{k+q}^b - \omega_0)}{( \omega^b_{k-q}
-\omega^a_{k-q}) (\omega^a_{k+q} + \omega^b_{k-q} - \omega_0 )} \Bigg],
\end{eqnarray}
}
\end{widetext}
where,
\begin{eqnarray}
\omega^{a,b}_k = \frac{\epsilon_k^a + \epsilon_k^b}{2} \pm \sqrt{\left(\frac{%
\epsilon_k^a - \epsilon_k^b}{2} \right)^2 + V_k^2}
\end{eqnarray}
are the energies of the hybrid bands. These bands are schematized in Fig. \ref{fig1}.

In the case hybridization vanishes, $%
V\rightarrow 0$, we get,
\begin{eqnarray}
\lim_{V\rightarrow0}\chi^{aa}_{0}(2q,\omega_0) = \sum_k \frac{1 -
f(\epsilon^a_{k+q}) - f(\epsilon^a_{k-q})}{\epsilon^a_{k+q} +
\epsilon^a_{k-q} - \omega_0},
\end{eqnarray}
and the condition $1-U\chi_{0}^{aa}(q=0,\omega_{0}=0)=0$, yields,
\[
\frac{1}{U}=\frac{\mathcal{V}}{(2 \pi)^{3}} \int d^{3}k \frac{\tanh(\beta
\varepsilon_{k}^{a} /2)}{2 \varepsilon_{k}^{a}},
\]
which is the BCS gap equation \cite{bcs}, where $\mathcal{V}$ is the volume and $\beta=1/k_B T$. At zero temperature this Thouless
condition is satisfied for any finite $U$ due to the logarithmic divergence
of the integral. On the other hand, in the presence of hybridization $V$,
for $q\rightarrow 0$ e $\omega_0 \rightarrow 0$ the susceptibility is given
by,
\begin{eqnarray}
\chi^{aa}_{0}\!(0,0)&\!=\! \sum_k\! \frac{1}{{(\omega^b_{k}}^2\!\!-\! {%
\omega^a_{k}}^2)} \left[ [1\!-\!2f(\omega^a_{k})]\frac{({\epsilon^b_{k}}%
^2\!\!-\!{\omega^a_{k}}^2)}{2\omega^a_{k}} \right.\nonumber\\&\!-\!\left. [1\!-\!2f(\omega^b_{k})]%
\frac{({\epsilon^b_{k}}^2\!-\! {\omega^b_{k}}^2)}{2\omega^b_{k}} \right].
\end{eqnarray}

In this case the condition $1-U\chi_{0}^{aa}(q=0,\omega_{0}=0)=0$ at $T=0$
yields a critical value for hybridization $V_c$ below which the system
becomes superconductor. Alternatively, for a given value of $V$, there is a
critical value of the attractive interaction above which the system is
superconductor. This condition turns out to be the same of that found
previously using a mean-field approximation. In this case, however, the
system starts in the superconducting phase and as hybridization reaches the
critical value $V_{c}$ it enters the normal phase continuously at a SQCP.
Further understanding of this instability is gained by examining Fig. \ref{fig1}. It shows the dispersion relations for the original bands ($V=0$) and when the hybridization is turned on. For simplicity we consider homothetic bands, $\epsilon_k^b=\alpha \epsilon_k^a$, ($\alpha>1$), with $\epsilon_k^a= k^2/2m_a -\mu$ ($\hbar=1$). The integration in $k$-space in the generalized susceptibility is done in a region of width $2k_C$ around the original Fermi wavevector $k_F$ ($ k_F^2/2m_a =\mu$). As V increases there is a reorganization of the electronic structure. The bands repel and the Fermi wave-vectors of the new hybrid bands eventually move out of the region of integration (see Fig. \ref{fig1}) avoiding the weak coupling logarithmic singularity. The interest of the model is that, even in the situation of Fig. \ref{fig1}, the system can sustain a superconducting ground state if the attractive interaction is sufficiently strong.
In these conditions, the equation $1-U\chi_{0}^{aa}(q=0,\omega_{0}=0)=0$ can be solved
numerically for finite temperatures and a critical line as a function of hybridization, $T_{c} \times
V$, ending at a SQCP can be obtained \cite{Padilha}. Close to $V_{C}$ the
critical temperature vanishes like, $T_{C} \propto \sqrt{V_{C}-V}$ with a
mean-field shift exponent $\psi=1/2$.

\begin{figure}[h]
\begin{center}
\includegraphics[width=1\linewidth,
  keepaspectratio]{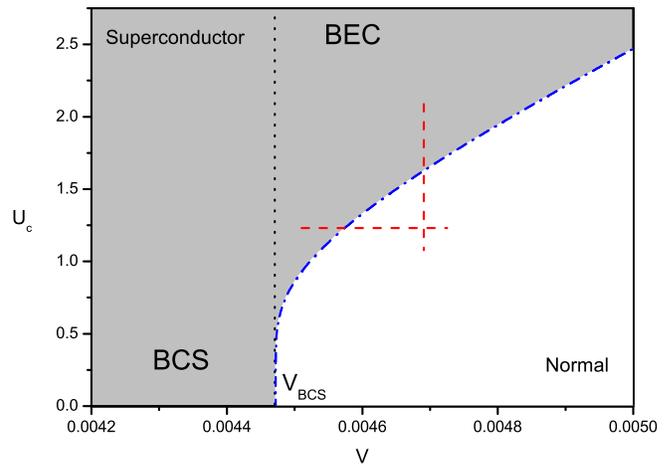}
\end{center}
\caption{(Color online) Zero temperature phase diagram for the intra-band superconductor for fixed chemical potential. In the normal phase, as the hybridization ($V$)
is reduced, for finite $U$, or $U$ is increased for  $V > V_{BCS}$  as indicated by the dashed lines, the normal metal becomes unstable to a
superconducting phase. This phase transition is in the universality class of the zero temperature Bose-Einstein condensation as discussed in the text. For $V \le V_{BCS}$, $T=0$, the system is a BCS superconductor for any value of $U \ne 0$. The values of the parameters used to obtain the curve are: $\alpha=5$ and $ \omega_c = v_F k_C=0.001$. All units are in terms of the bandwidth $\mu$.} \label{fig2}
\end{figure}

Solving the equation $1-U\chi_{0}^{aa}(q=0,\omega_{0}=0)=0$ for
$T=0$,  we obtain the zero temperature phase diagram of Fig.
\ref{fig2}. It shows the critical value of the interaction
\emph{versus} hybridization. Below a value that we called $V_{BCS}$
the system presents the usual BCS instability and is superconductor
for any finite $U$. For values of $V  >  V_{BCS}$,  a finite value
of the interaction is necessary for the existence of
superconductivity. So, in this case, we have a quantum phase
transition tuned by the strength of the hybridization for a given
value of interaction or tuned by the strength of the interaction for
a given value of hybridization. The value $V_{BCS}$ is that for
which the Fermi wave-vectors of the new hybrid bands coincide with
the limit of integration around the original Fermi wave-vector. For
homothetic bands $V_{BCS}=2 \sqrt{\alpha} v_F k_C$. As $V
\rightarrow V_{BCS}$ from above, the critical line vanishes as the
inverse of $\tanh^{-1}(V/V_C)$ (see dot-dashed line in Fig.
\ref{fig2}).

Further knowledge of the nature of the quantum phase transitions requires to analyze the
frequency and wave-vector dependence of the generalized susceptibility, Eq. \ref{ChiSim}. Here we are interested in the  metal-superconductor transition for $V >V_{BCS}$, i.e., in the case the system enters the superconducting state by reducing the hybridization or increasing $U$ as along the trajectories shown by the dashed lines of Fig. \ref{fig2}. In these conditions, we can obtain the real and imaginary part of the dynamic susceptibility. The imaginary part turns out to be zero and the real part is given by:
\begin{eqnarray}
Re[\chi^{aa}_0(2q,\omega_0 + i\eta)] &=
\frac{3}{16\pi^3}\frac{k_C}{k_F}\frac{1}{V}\left[1 +
\frac{\omega_0}{2\mu} \right.\nonumber\\&-\left. \frac{1}{4}\left(\frac{q}{k_F}\right)^2
\frac{k_F}{\delta k_{F}}F(\alpha) \right],
\end{eqnarray}
where $F(\alpha) = 10(1+\alpha)$, $\delta k_{F}=\mu/ \alpha v_F$ and $k_c$ the cut-off vector in k-space. Then, at the critical value of hybridization the system has propagating bosonic excitations with a quadratic dispersion relation given by:
\begin{eqnarray}
\omega_0=\frac{q^2}{2m_a}F(\alpha).
\end{eqnarray}

We can rewrite the dynamic susceptibility as:
\begin{eqnarray}
\chi^{aa}(2q,\omega_0)=\frac{\chi^{aa}_0(2q,\omega_0)}{V-V_C + \omega_0 - \mathcal{D}q^2}.
\end{eqnarray}

The critical line $V_{C}(U)$ or $U_{C}(V)$ for $V>V_{BCS}$ is shown
in Fig. \ref{fig2} (dot-dashed line). Along this line, the dynamic
critical exponent associated with the normal-superconductor
transition is $z=2$. This transition is in the same universality
class of the zero temperature Bose-Einstein condensation as a
function of the chemical potential \cite{muciobe}. This type of
transition occurs in antiferromagnetic systems at $T=0$ where
spin-waves can also present the phenomenon of Bose-Einstein
condensation as a function of the magnetic field \cite{muciobe}. It
is useful to express the results above using the language of the
renormalization group. The interaction $U$ is a relevant parameter
at the fixed point, $V_{BCS}$, $U=0$, and the quantum critical
behavior all along the critical line in Fig. \ref{fig2} for $V >
V_{BCS}$ is governed by a strong coupling fixed point  associated
with the Bose-Einstein condensation of Cooper pairs. Since the
effective dimension for this transition $d_{eff}=d+z$ is greater
than the upper critical dimension in three spacial dimensions, its
critical exponents assume Gaussian values. A scaling analysis
\cite{mac} using $z=2$ and Gaussian exponents allows to determine
the thermodynamical properties of the system in particular along the
critical trajectory, ($V=V_C$, $T \rightarrow 0$). The critical line in $d=3$ approaches the SQCP as \cite{mac},
$T_C \propto |V-V_C|^{2/3}$. The contribution of the critical fluctuations for the specific
heat along the critical trajectory \cite{mac} is given by
$C\propto T^{3/2}$. For disordered systems, the scattering of electrons by bosons with parabolic dispersion relation
yields a resistivity, $\rho \propto T$ and $\rho \propto T^{3/2}$ in $d=2$ and $d=3$ \cite{rivier}, respectively.

\subsection{Stability of the superconducting phase}

The superconducting phase we obtain for $V>V_{BCS}$ has interesting features. It has pairs of quasiparticles coexisting with gapless excitations at a Fermi surface. In some respects it resembles the breached pair superconducting phase proposed some time ago by Wilczek and Liu \cite{Wilczek}.
It is then natural to ask about the stability of the superconducting phase for large values of hybridization. For this purpose we study the difference between the ground state energy of the normal and superconducting phases as hybridization increases. The quasi-particle excitations in the superconducting phase have been obtained in Ref. \cite{Padilha} and are given by,
\begin{equation}
\Omega_k^{a,b}=\sqrt{A_{k}\pm\sqrt{B_{k}}}, \label{dispersion}%
\end{equation}
with,
\begin{equation}
A_{k}=\frac{\epsilon_{k}^{a2}+\epsilon_{k}^{b2}}{2}
+V^{2}+\frac{(\Delta^{aa})^{2}}{2}, \label{eq05}%
\end{equation}
and%
\begin{equation}
B_{k} =\left(  \frac{{\epsilon_{k}^{a}}^2-{\epsilon_{k}^{b}}^2+(\Delta^{aa})^2}{2}\right)
^{2}+V^{2}\left[\left(\epsilon_{k}^{a}+\epsilon_{k}^{b}\right)^{2}+(\Delta^{aa})^2\right],\label{eq06}
\end{equation}
where $\Delta^{aa}$ is the superconductor order parameter.
The free energy due to condensation is given by,
\begin{equation}\label{condensation}
\Delta E= -\int_{k_F-k_c}^{k_F+k_c} (-|\omega_k^b| -|\omega_k^a| +\Omega_k^{a}+\Omega_k^{b})d\mathbf{k} +\frac{ (\Delta^{aa})^2}{U}.
\end{equation}

\begin{figure}[h]
\begin{center}
\includegraphics[width=0.9\linewidth,
  keepaspectratio]{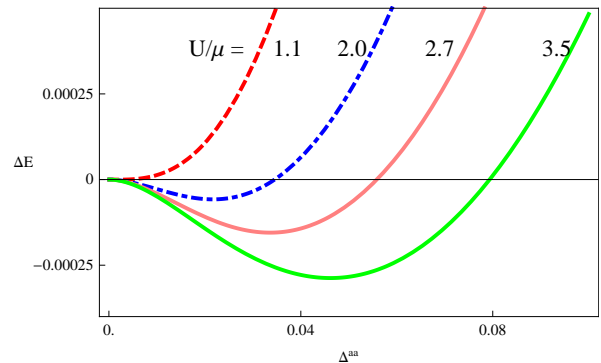}
\end{center}
\caption{(Color online) Condensation free energy, Eq. \ref{condensation}, for different values of the interaction. The system with a fixed value of hybridization ($V > V_{BCS}$) and  $U/\mu=1.1$ starts in the normal state with a minimum at $\Delta^{aa}=0$. As the attractive interaction $U/\mu$ increases, it becomes a superconductor through a continuous quantum phase transition which is in the universality class of the density-driven Bose-Einstein condensation. This phase transition corresponds to following the vertical dashed line in Fig. \ref{fig2}. Values of the parameters and units as in Fig. \ref{fig2}.} \label{fig3}
\end{figure}

This  vanishes in the normal state and its minimization with respect to the order parameter $\Delta^{aa}$ yields the gap equation for the intra-band superconductor order parameter.
In Fig. \ref{fig3} we show this energy  difference as the system crosses the phase transition line for a fixed $V > V_{BCS}$ with increasing $U$, as along the vertical dashed line in Fig. \ref{fig2}. The phase transition from the normal ($\Delta=0$) to the superconducting state with increasing $U$ is continuous or second order. The quantum phase transitions at the critical line $U_c(V)$ for $V > V_{BCS}$ as shown before are in the universality of the density-driven quantum Bose-Einstein condensation. The superconducting  phase is clearly associated with a minimum in the condensation energy as shown in Figure \ref{fig3}.

\begin{figure}[h]
\begin{center}
\includegraphics[width=0.9\linewidth,
keepaspectratio]{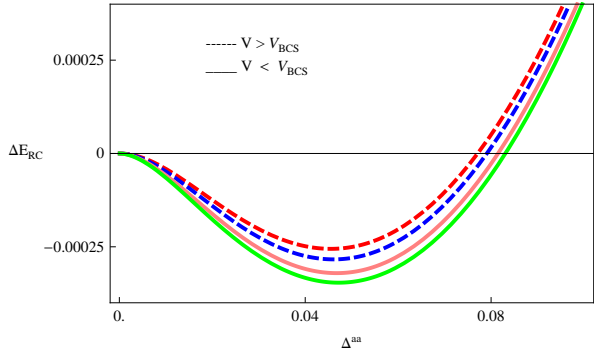}
\end{center}
\caption{(Color online) Condensation free energy, Eq. \ref{condensation}, as the system moves from the BEC state to the BCS state with reducing hybridization (horizontal dashed line in Fig. \ref{fig2}). As shown, both phases are smoothly connected.  The superconductivity minimum just shifts and gets deeper continuously as $V$ decreases passing through $V_{BCS}$. Values of the parameters and units as in Fig. \ref{fig2}. } \label{fig4}
\end{figure}

Figure \ref{fig4} shows the energy difference $\Delta E$ as the superconductor moves from the Bose-Einstein condensation regime at $V > V_{BCS}$ to the BCS regime for $V< V_{BCS}$ with decreasing hybridization. It is clear from the figure that this change of regime has a smooth behavior and nothing special occurs at $V_{BCS}$ for a finite value of $U$. As the BCS phase is stable for $V < V_{BCS}$, this figure with the continuous behavior of the ground state along the change of regime implies also the stability of the BEC-like phase for $V > V_{BCS}$.

\begin{figure}[h]
\begin{center}
\includegraphics[width=0.9\linewidth,
keepaspectratio]{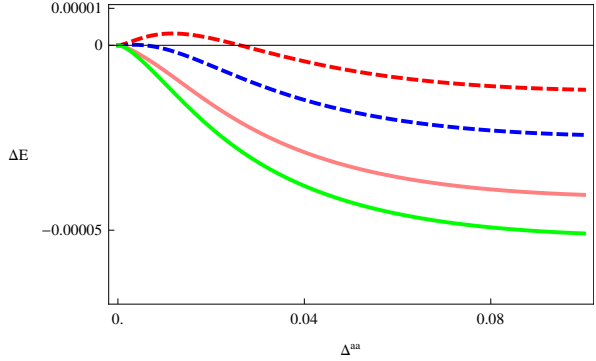}
\end{center}
\caption{(Color online) Difference of the energies of the superconducting states integrating in  shells of width $k_c$ around $k_F$ and $k_{FB}$ (see Fig.\ref{fig1}), for the same values of $V$ and parameters as in Fig.\ref{fig4}. Dashed lines are for $V>V_{BCS}$ and continuous lines for $V<V_{BCS}$. Notice that the BCS state for $V<V_{BCS}$ has always a lower energy when integrating around the original $k_F$ ($\Delta E < 0$). Also for $V>V_{BCS}$, $\Delta E < 0$ at the minimum of the superconducting state showing the stability of the breached pair-like state for $U>U_{c}(V)$ with  respect to the BCS state obtained integrating around $k_{FB}$. } \label{fig5}
\end{figure}

Our results are obtained with the superconducting pairs being formed close to the original Fermi wave-vectors
of the unhybridized bands.  One may argue that as hybridization is turned on, we should integrate not around the original $k_F$ but around $k_{FB}$ of the new hybrid band in which case mixing would never suppress superconductivity. However, in the process of diagonalization  the two-band problem, terms of the type $\alpha^{\dagger} \alpha^{\dagger}$, $\alpha^{\dagger} \beta^{\dagger}$ and $\beta^{\dagger} \beta^{\dagger}$ are generated. The operators $\alpha^{\dagger}$, $\alpha$, $\beta^{\dagger}$, $\beta$ are the creation and annihilation operators of the new hybrid bands. Then all types of superconducting correlations among the hybrid bands appear and there is no particular reason to integrate around, say $k_{FB}$. In order to expose the complexity of this problem, we  calculate the difference between the ground state free energies of the superconducting states obtained integrating around $k_F$ and $k_{FB}$. This difference is shown in Fig. \ref{fig5}. For the BCS type states with $V<V_{BCS}$, this is always negative showing that integration around $k_F$ leads to a lower energy. For $V > V_{BCS}$, the breached pair type of superconducting state  has, at the value of $\Delta^{aa}$ for which the energy is a minimum, a lower energy than that of the BCS state obtained integrating around $k_{FB}$ for $V<V_c(U)$.

\section{The BCS-BEC Crossover}

From Fig. \ref{fig2} one can see that for values of $V$ slightly
greater than $V_{BCS}$ the value of $U_c$ is already around $E_F$,
so is seems important the study of the transition in a formalism in
which we are able to reach also the strong coupling limit. In this
section we study the quantum phase transition tuned by hybridization
in a range of the interaction strength. Previous studies in one-band
systems have shown that the evolution between these two different
regimes is continuous \cite{SadeMelo1,Nozieres,PhysToday}. For the
case of a two-band system with d-wave symmentry it was also found
that this crossover is smooth \cite{Dinola}.

We start with the generalized gap equation to describe the weak to strong coupling crossover extended for the  two-band case with hybridization, $\frac{1}{U}=\chi_{0}^{aa}(2q,\omega_{0})$ In this equation  $\chi_{0}^{aa}(2q,\omega_{0})$ is given by Eq. \ref{ChiSim} with the $\omega^{a,b}_k$ replaced by the dispersion relations in the superconducting state,   Eqs. \ref{dispersion}-\ref{eq06}.

In this case a BCS-type cut-off cannot be used to treat the strong coupling limit, in which all the particles must interact, and not just those around the Fermi surface as in the BCS limit. We thus introduce the concept of \emph{s}-wave scattering length, $a_s$, to regulate the ultraviolet divergence in the gap equation that arises due to the sum over all energies \cite{SadeMelo2}. It is related to the coupling strength $U$ through:
\begin{equation}
\frac{m}{4\pi a_s}=-\frac{1}{U} + \sum_k \frac{1}{2(\epsilon_k^b + \mu)},
\end{equation}
and can be positive or negative. When $1/k_Fa_s \rightarrow -\infty$ one obtains the weak coupling regime, while $1/k_Fa_s \rightarrow \infty$ gives the strong coupling limit.

Eliminating the coupling U from the gap equation using the relation above, we find:
\begin{equation}\label {eq25}
-\frac{m}{4\pi a_s}=\chi_{0}^{aa}(2q,\omega_{0}) - \sum_k \frac{1}{2(\epsilon_k^b + \mu)}.
\end{equation}

For the purpose of solving this equation and determine $V_c$ we can
fix the chemical potential or the total number of particles. For the
second case we need first to obtain the chemical potential as
function of the new coupling parameter $1/k_Fa_s$. From the
propagators in the superconducting phase given in \cite{Padilha} we
can find the number equation, that guarantees the conservation of
the total number of particles through the variation of the chemical
potential,
\begin{eqnarray}\label {eq26}
N=\sum_k \left\{1-\frac{1}{2({\Omega_k^a}^2-{\Omega_k^b}^2)}\left[\frac{(\epsilon_k^a+ \epsilon_k^b)({\Omega_k^a}^2+V^2)}{\Omega_k^a}\right.\right.\nonumber\\ \left.\left.-\frac{\epsilon_k^a[\Delta^2+\epsilon_k^b(\epsilon_k^a+ \epsilon_k^b)]}{\Omega_k^a}-\frac{(\epsilon_k^a+ \epsilon_k^b)({\Omega_k^b}^2+V^2)}{\Omega_k^b}\right.\right.\nonumber\\ \left.\left.+\frac{\epsilon_k^a[\Delta^2+\epsilon_k^b(\epsilon_k^a+ \epsilon_k^b)]}{\Omega_k^b}\right]\right\}.
\end{eqnarray}

\begin{figure}[h]
\begin{center}
\includegraphics[width=0.9\linewidth,
keepaspectratio]{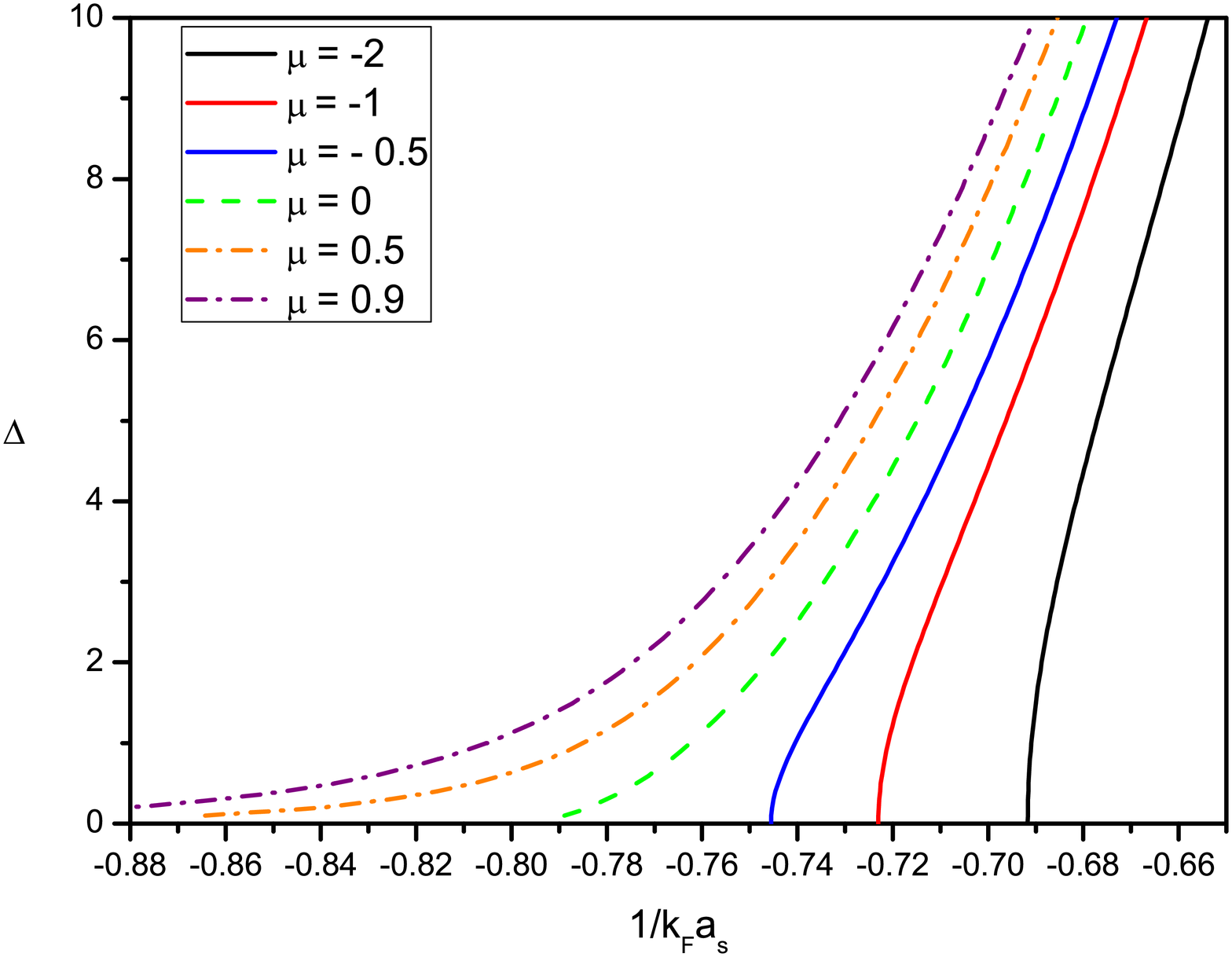}
\end{center}
\caption{(Color online) Zero temperature gap $\Delta$ as function of
the interaction parameter $1/k_Fa_s$, for the mass ratio $\alpha=5$
and hybridization V=0.01 for different values of fixed chemical
potential.} \label{fig6}
\end{figure}

\begin{figure}[h]
\begin{center}
\includegraphics[width=1\linewidth,
keepaspectratio]{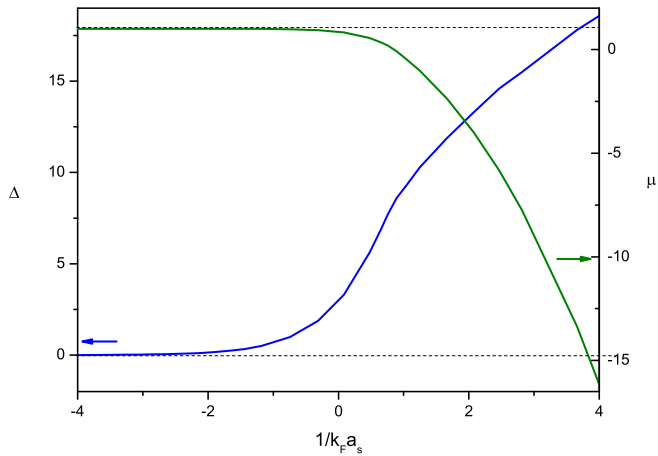}
\end{center}
\caption{(Color online) Zero temperature gap $\Delta$ and chemical
potential $\mu$ as functions of the interaction parameter
$1/k_Fa_s$, for the mass ratio $\alpha=5$ and hybridization V=0.01.}
\label{fig7}
\end{figure}

For fixed chemical potential we can find $\Delta(=\Delta^{aa})$ versus $1/k_Fa_s$
for different values of $\mu$, as shown in Fig. \ref{fig6}. For
fixed number of particles equations \ref{eq25} and \ref{eq26} must
be solved self-consistently. Fig. \ref{fig7} shows the behavior of
the superconductor order parameter and the chemical potential as
function of the coupling parameter at $T=0$. One can se that in the
BCS limit the chemical potential practically does not differ from
the Fermi energy, and the superconducting gap is much smaller than
$E_F$. Increasing the coupling, the pairs become more tightly bound
and the chemical potential decreases and becomes negative.

\begin{figure}[h]
\begin{center}
\includegraphics[width=1\linewidth,
keepaspectratio]{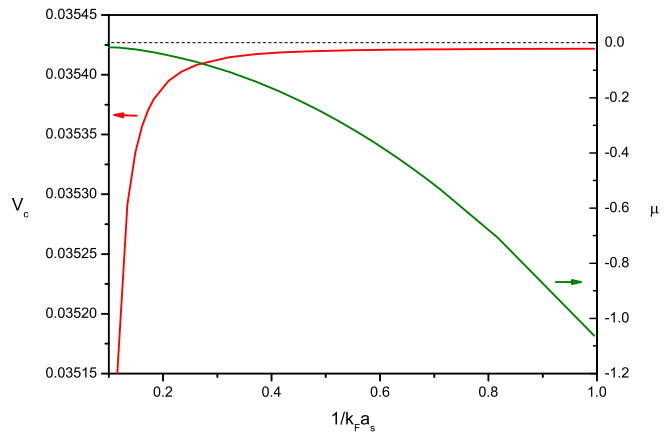}
\end{center}
\caption{(Color online) Critical values of the hybridization $V_c$
and chemical potential at $T=0$ as functions of the coupling
strength $1/k_Fa_s$, for the mass ratio $\alpha=5$.} \label{fig8}
\end{figure}

The behavior of the critical value of the hybridization as function
of $1/k_Fa_s$ is shown in Fig. \ref{fig8}.   An increase in
hybridization for a fixed value of the interaction suppresses
superconductivity continuously,  as can be seen in Fig.\ref{fig4}.
The critical value of hybridization increases with the coupling
parameter and tends to a saturation value. Notice that in this case
there is a critical value of hybridization only for strong couplings
as the chemical potential becomes negative. It can be understood
analyzing Fig. \ref{fig6}, in which we find that the order parameter
goes to zero for a definite value of $1/k_Fa_s$ just for negative
values of the chemical potential (solid lines).

\section{Uemura relation and scaling close to the SQCP}

Close to a superconducting quantum critical point, the vanishing of the superfluid density at $T=0$ is governed by the scaling relation \cite{mac}, $\rho_S(T=0) \propto |g|^{2-\alpha-2\nu}$ where $\alpha$ and $\nu$  are  critical exponents  associated with the SQCP and $|g|=|V-V_C|$ measures the distance to the transition.

We are interested in dimensions  $d\ge2$, such that, for the  superconducting quantum phase transition considered here which is in the universality class of the density-driven Bose-Einstein condensation with dynamic exponent $z=2$,  the effective dimension $d_{eff}=d+z \ge =4$. This coincides or is larger than the upper critical dimension $d_c=4$ for this transition \cite{muciobe}. In this case using the mean-field values, $\alpha=0$ and $\nu=1/2$, we find that $\rho_S$ scales linearly with the distance to the SQCP, $\rho_S(T=0) \propto |g|$ as $g \rightarrow 0$.

On the other hand close to the SQCP, as $|g| \rightarrow 0$, the superconducting critical temperature vanishes with the distance to the SQCP as, $T_c \propto |g|^{\psi}$ where $\psi$ is the {\it shift exponent}. For $d+z\ge d_c=4$, the interaction $u$ between the bosons is dangerously irrelevant and the shift exponent is determined by the dynamic exponent and the dimension of the system \cite{muciobe,mac,millis}, $\psi=z/(d+z-2)$ \cite{chakra}.

Finally, we can derive a Uemura relation \cite{uemura} relating the $T=0$ superfluid density and the critical temperature for close to a SQCP. Due to the proximity to the SQCP this is completely determined by scaling. For  $d+z \ge d_c$ this is given by, $T_c \propto \rho_S^{\psi}$ with $\psi=z/(d+z-2)$. In our case with $z=2$, we find, $T_c \propto \rho_S$ in $d=2$ and $T_c \propto \rho_S^{2/3}$ in $d=3$.

\section{Conclusions}

We have studied the normal-superconductor quantum phase transition induced by hybridization in a two-band system using a new method recently introduced to deal with systems which are coupled
to a space and time dependent perturbation. The results obtained
are valid to first order in the perturbation and coincide with those
of linear response theory. However, in our approach the basic
elements to be calculated are single particle Green's functions and not the usual two
particle propagators of linear response theory.
In the case of superconductivity, to calculate the superconducting  response, we have to introduce a fictitious field which
couples to the superconductor order parameter. Starting from the
normal phase, we obtained a generalized wave-vector and frequency
dependent susceptibility. In the static and homogeneous limit, we
argued that the divergence of this susceptibility signals the
instability of the system to a  superconducting state. For the standard BCS problem with a single attractive band, this condition
is similar to the Thouless criterion for BCS
superconductivity. For multi-band superconductors with
intra-band interactions, we had obtained previously  using a mean-field BCS
approximation that superconductivity can be suppressed by
increasing hybridization \cite{Padilha}. Here we further clarify the mechanism of suppression and extend the  mean-field approach to include fluctuations close to the quantum superconductor-normal phase transition. We find that for finite attractive interactions, as hybridization increases, the BCS superconductor has a smooth crossover to a superconducting breached pair-like state before becoming normal. The quantum superconductor-normal phase transition is in the universality class of the density-driven Bose-Einstein condensation.

We have also investigated the case of fixed number of particles in the presence of hybridization. We have determined the $T=0$ phase diagram of the critical value of the hybridization {\it versus}  the strength of the interaction given in terms of the scattering length. In this case superconductivity is also suppressed with the increase of hybridization, but only in the strong coupling limit for which the chemical potential becomes negative.

\begin{acknowledgements}
We would like to thank Igor Padilha and Catherine Pepin for useful discussions. We also thank the Brazilian agencies, CNPq and FAPERJ for partial financial support.
\end{acknowledgements}

\end{document}